\documentclass[Print]{Style/isecure-v24}

\usepackage{lipsum}
\usepackage[colorlinks, bookmarksnumbered, linkcolor=darkblue, citecolor=darkred, urlcolor=darkgreen]{hyperref}
\usepackage[numbers,sort&compress]{natbib}
\usepackage[all]{hypcap}
\usepackage{Style/picins}
\usepackage{lettrine}
\usepackage{wrapfig}
\usepackage{graphicx}
\usepackage{subcaption}
\usepackage{amsthm}
\usepackage{amssymb,amsmath,amsfonts,eqnarray,breqn}
\usepackage{fancyhdr}
\usepackage{ragged2e}
\usepackage{enumitem}   
\usepackage{refcount}
\usepackage[protrusion=true,expansion=true]{microtype}
\usepackage{float}
\usepackage[greek,english]{babel}
\usepackage[LGR,T1]{fontenc}
\usepackage[T1]{fontenc}
\usepackage{lmodern}
\usepackage{url}
\usepackage{hyperref }
\usepackage{natbib}
\usepackage{tabulary,rotating}
\usepackage{longtable}
\usepackage{xtab,booktabs}
\usepackage{tabularray}

\DisableLigatures[f]{encoding=*,family=*}
\graphicspath{{Images/},{Style/}}
\DeclareGraphicsExtensions{.jpg,.pdf,.png}
\input{Style/Theorem-Styles.tex}
\articletype{Short Paper}

\usepackage{titlesec}
\titleformat{\paragraph}
{\normalfont\normalsize\bfseries}{\theparagraph}{1em}{}
\titlespacing*{\paragraph}
{0pt}{3.25ex plus 1ex minus .2ex}{1.5ex plus .2ex}

\journal{ISeCure}
\company{ISC}
\copyrightyear{2023}
\journalsite{http://www.isecure-journal.org}
\firstpage{1} 

\makeatletter
\begingroup \lccode`+=32 \lowercase
 {\endgroup \def\Url@ObeySp{\Url@Edit\Url@String{ }{+}}}
 \def\Url@space{\penalty\Url@sppen\ }
\makeatother

\usepackage[displaymath]{lineno}  
\newcommand*\patchAmsMathEnvironmentForLineno[1]{
  \expandafter\let\csname old#1\expandafter\endcsname\csname #1\endcsname
  \expandafter\let\csname oldend#1\expandafter\endcsname\csname end#1\endcsname
   \renewenvironment{#1}
     {\linenomath\csname old#1\endcsname}
     {\csname oldend#1\endcsname\endlinenomath}}
\newcommand*\patchBothAmsMathEnvironmentsForLineno[1]{
  \patchAmsMathEnvironmentForLineno{#1}
  \patchAmsMathEnvironmentForLineno{#1*}}
\AtBeginDocument{
\patchBothAmsMathEnvironmentsForLineno{equation}
\patchBothAmsMathEnvironmentsForLineno{align}
\patchBothAmsMathEnvironmentsForLineno{flalign}
\patchBothAmsMathEnvironmentsForLineno{alignat}
\patchBothAmsMathEnvironmentsForLineno{gather}
\patchBothAmsMathEnvironmentsForLineno{multline}
}

\begin{document}
\begin{frontmatter}


\def\NoDingTitle{Using ChatGPT as a Static Application Security Testing Tool}
\title{\NoDingTitle\textsuperscript{ }}


\author[a1,CorAuth]{Atieh Bakhshandeh}, 
\ead{ bakhshandeh@rcdat.com}
\author[a1]{Abdalsamad Keramatfar},
\ead{keramatfar@rcdat.ir}
\author[a1]{Amir Norouzi}, and%
\ead{norouzi@rcdat.ac.ir}
\author[a1]{Mohammad Mahdi Chekidehkhoun}
\ead{chekidekhoon@rcdat.ir}

\address[a1]{Research Center for Development of Advanced Technologies, Tehran, Iran}

\corauth[CorAuth]{Corresponding author.}

\begin{abstract}
In recent years, artificial intelligence has had a conspicuous growth in almost every aspect of  life. One of the most applicable areas is security code review,  in which a lot of AI-based tools and approaches have been proposed. Recently, ChatGPT has caught a huge amount of attention with its remarkable performance in following instructions and providing a detailed response. Regarding the similarities between natural language and code, in this paper, we study the feasibility of using ChatGPT for vulnerability detection in Python source code. Toward this goal, we feed an appropriate prompt along with vulnerable data to ChatGPT and compare its results on two datasets with the results of three widely used Static Application Security Testing tools (Bandit, Semgrep and SonarQube). We implement different kinds of experiments with ChatGPT and the results indicate that ChatGPT reduces the false positive and false negative rates and has the potential to be used for Python source code vulnerability detection.
\end{abstract}


\makeatother

\end{frontmatter}

\addtolength{\parskip}{2mm}


\section{Introduction} \label{sec:intro}
 
Today, almost all technologies are strongly dependent on source code. Therefore, code is of increasing importance. A glance at the number of lines of used codes in some well-known tools is evidence for this claim. For instance, the number of GitHub repositories increased from 100 million in 2018 to 200 million in 2022 \cite{githubstat}. It is clear that the increase in the amount of code will  lead to more security requirements in programming. MITRE\footnote{Massachusetts Institute of Technology Research and Engineering} and other studies indicate the growth in the number of vulnerabilities in recent years~\cite{mitre,vulIncrease1,vulIncrease2,vulIncrease3,vulIncrease4,vulIncrease5,vulIncrease6}. Specifically, software vulnerability is of great importance. Software vulnerability is a technical vulnerability that can be used for violating its security policies. Such vulnerabilities can be exploited which in turn leads to data leakage and tampering and even denial of services. 

Static source code analysis is a method for finding code vulnerabilities that is done by automatically examining the source code without having to execute the program. Static Application Security Testing (SAST) tools analyze a piece of code or a compiled version of it in order to identify its security problems. Covering a wide range of errors and  high accuracy are two important features of SAST tools \cite{sastProperties}. Most of the well-known SAST tools such as Semgrep, Bandit, and SonarQube often use rule-based techniques to find the vulnerable patterns of code. However, these tools have shown to have their own flaws, including a high rate of false positives and false negatives \cite{vulIncrease2}. The more false positives a SAST tool returns, the more time and effort are required by a security expert to validate the findings of the SAST tool. Moreover, this will increase the error rate by humans, which may then lead to ignoring some vulnerabilities. On the other hand, a high rate of false negative leads to catastrophic events.

In recent years, Machine Learning (ML) and deep learning have had remarkable advances in various areas such as natural language processing \cite{Machine,keramatfar2022graph}. Therefore, considering the high similarity between code and natural languages, the deep learning-based models are expected to be successful in code processing tasks. Likewise, studies in this area have shown the interest of researchers in using deep learning techniques in vulnerability detection \cite{deeplearning1,deeplearning2}. Machine learning models can automatically learn the patterns of software vulnerabilities based on datasets. Furthermore, research indicates that ML models have fewer false positives compared to SAST tools \cite{vulIncrease4,deeplearning6}. Results of a new research have shown the superior performance of deep learning-based models over three open-source tools in C/C++, reducing false positives and negatives rate at the same time \cite{LR1}.

Recently, ChatGPT, an AI-powered chatbot tool that uses Natural Language Processing (NLP) and machine learning algorithms to understand and respond to customer inquiries, has drawn a lot of attention. ChatGPT is vital for business professionals for several reasons. It can help save time and resources by automating tasks requiring human intervention. An important point to note is that, ChatGPT has been trained with a huge amount of data till 2021 so that it can be a great help in finding known patterns in thousands of packages in automated way. The model is also trained on a large amount of code and is thus able recognize common patterns. In this paper, we evaluate the performance of ChatGPT in identifying security vulnerabilities of Python codes and compare the results with three well-known SAST tools for Python vulnerability detection ( Bandit, Semgrep and SonarQube). The reason for choosing the Python language is that  in 2022, Python was known to be the most popular programming languages along with Java, based on the Popularity of Programming Language Index (PYPL) and IEEE reports. Also, Stackscale ranked Python at the third place \cite{spectrum}. Although Python is mainly used in the scope of machine learning and data science, its applications are not only limited to these fields and with its famous frameworks such as Django and Flask, it is prone to vulnerabilities. The rest of the paper is organized as follows: In Section \ref{sec:related}, we provide a brief literature review of this area. Section \ref{sec:dataset} is dedicated to datasets we used. Section \ref{sec:modelAPI} provides the details of the experiments we performed with ChatGPT. In Section \ref{sec:evaluation}, we present the evaluation and analysis of the obtained results\footnote{ https://github.com/abakhshandeh/ChatGPTasSAST.git}. In Section \ref{threats_section} we discuss some factors that may threaten the validity of the results. Finally, Section \ref{sec:conclusion} concludes the paper.


\section{Related Work} \label{sec:related}
In this section, we review some of the works that  used different kinds of AI models for vulnerability detection. Note that we did not focus on works which proposed models for repairing the identified vulnerabilities. When it comes to artificial intelligence, the main idea is the use of supervised learning. Therefore, various machine learning models used methods of feature engineering such as the number of lines of the code, code complexity and the number of operations and also utilized textual features\cite{LR1,LR2}. In general, research shows that text-based models have better performance over feature engineering and the studies also admit that machine learning models outperform the existing SAST tools.

Recently, more research has been devoted to deep learning. In this scope, researchers often used different deep learning models such as Convolutional Neural Network (CNN), Long Short-Term Memory (LSTM), and Multilayer Perceptron (MLP) \cite{deeplearning2,deeplearning3,deeplearning4,deeplearning5}. Some of the models were based on different kinds of code property graphs and used Graph Neural Network \cite{deeplearning2,deeplearning6}, while some others only relied on tokens \cite{deeplearning5}. A new study has investigated the way deep learning models function in vulnerability detection tasks. The results of this study reveal some points: first, the results of different models are not compatible with each other. Second, fine-tuned models have shown better performance in this field. Third, usually 1000 samples of each class are enough for the training of a neural network and finally, models usually use the same features for prediction \cite{deeplearning7}.  Although  studies approved the superiority of graph-based models, a new study indicates the superior performance of transformer-based models over graph-based ones \cite{deeplearning8}. In 2022, Hanif and Maffeis proposed a model named VulBERTa \cite{deeplearning9}. This model is based on RoBERTa and is used for vulnerability detection in C/C++ codes. Another recent study also has used BERT architecture and CodeBERT vectors for predicting code vulnerabilities. The results of this study approve the superiority of transformer-based models over traditional deep learning models and also graph-based models \cite{deeplearningg1}. Overall, it seems that transformer-based models are effective in this area. Another recent work has evaluated ChatGPT as a large language model for detecting vulnerabilities in Java source codes and compared the results with a dummy classifier and achieved no better results than it \cite{deeplearningg2}. However, there is still no academic study about comparing the results of the ChatGPT model with traditional SAST tools for Python, and this paper aims to answer the question of whether the ChatGPT model is outperforming SAST tools or not?


\section{Datasets Description } \label{sec:dataset}
In this section, we provide the details about our dataset and the labels we used. Our dataset consists of 156 Python code files. These files contain 130 files of the securityEval dataset which is proposed in \cite{siddiq}. As the authors mentioned, these 130 files cover 75 vulnerability types that are mapped to Common Weakness Enumeration (CWE). The remaining 26 files belong to a project called PyT in which the author developed a tool for Python code vulnerability detection and used these 26 vulnerable code files for evaluating his tool \cite{pyt1,pyt2}. Since the used datasets do not provide the specific line of vulnerability, a security expert of our team rechecked the data and specified the vulnerable line. We identified the corresponding line of code of  CWEs that were assigned in the labels of these files with the help of a security expert. The datasets' information and the distribution of their corresponding labels are presented in \autoref{dataset_table} in appendix \ref{sec:appendix}.
%

\section{Working with ChatGPT API} \label{sec:modelAPI}
In this section, we provide the details of the process of utilizing the ChatGPT model API for identifying vulnerabilities. In this study, we used the GPT-3.5-turbo model. The GPT-3.5-Turbo model can accept a series of messages as input, unlike the previous version that only allowed a single text prompt. This capability provides some interesting features, such as the ability to store prior responses or query with a predefined set of instructions with context. This is likely to improve the generated response. The GPT-3.5-Turbo model is a superior option compared to the GPT-3 model, as it offers better performance across all aspects while being 10 times more cost-effective per token. We did four kinds of experiments using the GPT-3.5-Turbo model.

\begin{enumerate}
\item In our first experiment, we give the model the vulnerable files and ask it whether they contain any security vulnerabilities or not, without specifying the corresponding CWEs. We ask the model to just return the line number of the vulnerability if it contains any. Then, we compare these lines with ground truth labels. In effect, this experiment is a binary classification.

\item In our second experiment, we provide the list of the corresponding CWEs and ask the model to find the vulnerabilities from the labels' list in the Python vulnerable file. In this experiment, we ask the model to respond in JSON format like [\{"label": "CWE-X", "line of Code": "line no."\}] so we can compare our results with those of SAST tools.

\item In the third experiment, for each of vulnerable files, we give the model all the labels returned from Bandit, Semgrep and SonarQube tools for the Python code, as the classes that ChatGPT should use. We then ask the model whether each vulnerable file contains any of those vulnerabilities or not? Here, the main difference with our second experiment is that we specify the classes per vulnerable file separately. In other words, we use the model as an assistant for the SAST tools to verify the detected vulnerabilities by them. In this experiment, we use the same JSON format as the second experiment for the responses. Note that in this experiment, although we provide the labels' list beforehand for each vulnerable file, in some cases the model has returned a new CWE which is not among its input labels. This is a natural behavior seen from a language model and in order to address this issue in our evaluation, we consider two cases: In one case, we ignore the new labels and calculate the metrics without considering them. This policy can reduce the number of false positives of SAST tools. In another case, we consider them as well and this time the number of false negatives may decrease.

\item In our fourth experiment, we do not provide any label list for the model and ask it to detect the vulnerabilities in the files and determine their corresponding CWEs from its own trained knowledge. Here, the format of the responses is the same JSON structure in the previous experiments.
\end{enumerate}
 
To use the model for our experiments, we put all the vulnerable python codes of our dataset in a directory and called GPT-3.5 API with an optimized prompt for each of the vulnerable python files. The choice of prompt is the most challenging task in this process, as it has direct effect on the results the model provides. We optimize our prompts according to \cite{chatgptcourse}. \autoref{prompt_table} provides the list of prompts we used for each experiment.
\subsection{ Parameters}\label{parameters_section}
The parameters of the experiment are the prompt which contains the instructions the model will execute and a parameter called temperature that determines the randomness level of the model response. The temperature can take the values between 0 to 6 with 6 giving the most random output. Because we want to specify labels for a piece of code, the nature of our job is deterministic, so we set the temperature to zero to get the most stable behavior. Another important parameter is the prompt which is very influential in the results and adjusting it to get the best results is a challenging task. We used the prompts given in ~\autoref{prompt_table}. We chose our prompts based on \cite{chatgptcourse}. According to \cite{chatgptcourse}, some of the key points to create an optimized prompt are: using delimiters such as triple quotes or triple backticks to specify the piece of code, asking for structured output such as JSON, HTML, etc., specifying the steps to complete a task in a clear way, instructing the model to work out its own solutions before rushing to a conclusion. In order to show the sensitivity of ChatGPT to its prompts, an example is provided in \autoref{prompt_change_comparison_table} where a prompt with minor modification is given to the model and the model responses with different answers in which the second response contains one less vulnerability compared to the first one.

\begin{table}
	\centering
	\caption{Comparison of the responses of two Prompts with minor change}\label{prompt_change_comparison_table}
	\begin{tabular}{p{0.5\linewidth} | p{0.5\linewidth}}
		\cline{1-2} 
		Prompt~~~~~ & Response~~~~  \\ 
		\hline 
		find all the vulnerabilities with the CWE standard in the python code which
		is delimited with triple backticks. also give the line of the vulnerability in the code.
		python code:$'''$\{Vul_code\}$'''$
		Format your response as a list of JSON objects with \
		$"$label$"$ and $"$line of Code$"$ as the keys for each element.
		only answer with JSON.& [
  		\{
		    $"$label$"$: $"$CWE-78: Improper Neutralization of Special Elements used in an OS Command ($'$OS Command Injection$'$)$"$,
		    $"$line of Code$"$: $"$command = $'$echo$'$ + param + $'$ >> $'$ + $'$menu.tx$t'"$
		 \},
		  \{
		    $"$label$"$: $"$CWE-78: Improper Neutralization of Special Elements used in an OS Command ($'$OS Command Injection$'$)$"$,
		    $"$line of Code$"$: $"$subprocess.call(command, shell=True)$"$
		\},
		\{
		   $"$label$"$: $"$CWE-78: Improper Neutralization of Special Elements used in an OS Command ($'$OS Command Injection$'$)$"$,
		    $"$line of Code$"$: $"$subprocess.call($'$echo Menu: > menu.txt$'$, shell=True)$"$
		 \}
		]\\
		\hline
		find all the vulnerabilities with the CWE standard in the python code. also give the line of the vulnerability in the code.
		python code: $'''$\{Vul_code\}$'''$
		Format your response as a list of JSON objects with \
		$"$label$"$ and $"$line of Code$"$ as the keys for each element.
		only answer with JSON.&[
		 \{
		    $"$label$"$: $"$Command Injection$"$,
		    $"$line of Code$"$: $"$subprocess.call(command, shell=True)$"$
		  \},
		  \{
		    $"$label$"$: $"$Command Injection$"$,
		    $"$line of Code$"$: $"$subprocess.call($'$echo Menu: > menu.txt$'$, shell=True)$"$
		  \}
		]\\
		\hline		
	\end{tabular} 
\end{table}

\begin{table}

	\centering
	\caption{GPT-3.5  prompts used. Vul_code refers to the vulnerable code and labels1 is the list of labels of all vulnerable files and labels2  is the labels of each vulnerable file that is iterated through a loop.}\label{prompt_table}
	\begin{tabular}{p{0.35\linewidth} | p{0.6\linewidth}}
		\cline{1-2} 
		Experiment No.~~~~~ & Prompt~~~~  \\ 
		\hline 
		$Experiment 1$ & You will be provided with a python code delimited by triple backticks. 
          		  If it contains any security vulnerability, identify the lines of vulnerable code and only write the line in quotation\ 
            	  If the code does not contain a vulnerability,\ 
           	 then simply write None.
        		 python code: $'''$\{Vul_code\}$'''$ \\ 
		\hline
		$Experiment 2$ & Which of the following vulnerabilities from list of vulnerabilities exist in the python code which
           	 is delimited with triple backticks. also give the line of the vulnerability in the code. 
		 python code:$'''$\{Vul_code\}$'''$
		 list of vulnerabilities: \{$"$, $"$.join(labels1)\}
           	 Format your response as a list of JSON objects with \
            	$"$label$"$ and $"$line of Code$"$ as the keys for each element.
           	 only answer with JSON.  \\ 	
		\hline	 	
		$Experiment 3$ & Which of the following vulnerabilities from list of vulnerabilities exist in the python code which
            	is delimited with triple backticks. also give the line of the vulnerability in the code.
           	python code: $'''$\{Vul_code\}$'''$
           	list of vulnerabilities: \{$"$, $"$.join((labels2)\}
           	Format your response as a list of JSON objects with \
          		$"$label$"$ and $"$line of Code$"$ as the keys for each element.
            	only answer with JSON.  \\ 		 
		\hline 
		$Experiment 4$ & Your task is to determine whether the following python code which is delimited with triple backticks,is vulnerable or not?
         		   identify the following items:
                		- CWE of  its vulnerabilities. \
                		- lines of vulnerable code. \
	            Format your response as a list of JSON objects with \
	           $"$label$"$ and $"$line of Code$"$ as the keys for each vulnerability.
	            If the information isn't present, use $"$unknown$"$ \
	            as the value.
	            Make your response as short as possible and only answer with JSON.
	            python code:$'''$\{Vul_code\}$'''$  \\ 		 
		\hline
		
	\end{tabular} 
\end{table}

\section{ Results} \label{sec:evaluation}
In this section, we provide the results of our experiments. First we explain the metrics we use for evaluating our work and then we present  GPT-3.5 results and compare them with three popular SAST tools for Python vulnerability detection. To be more precise, we perform the following actions: We give a dataset of 156 vulnerable python codes to Bandit, Semgrep and SonarQube SAST tools and we also query the ChatGPT model with our dataset using the appropriate prompts. We then calculate the following metrics for each of the tools' results and the model result based on our ground truth labels. Finally, we compare the results of the tools with GPT-3.5 model. 
\subsection{Evaluation Metrics}
In classification, we have condition positive which indicates the number of real positive cases in the data. Similarly, there is a condition negative which is the number of real negative cases in the data. Based on these conditions, there will be four parameters: true positive (TP) which is the number of positive examples labeled as such, true negative (TN) that is the number of negative examples labeled as such, false positive (FP) that is the number of negative examples labeled as positive and false negative (FN) that is the number of positive examples labeled as negative. We define precision, recall and F-measure according to the following formulas \cite{bakhshandeh}.
\begin{itemize}
\item Precision: It answers the question that out of all the examples the classifier labeled positive, what proportion were correct? It is defined according to the following equation:
\begin{equation}
	precision=\frac{TP}{TP+FP}
\end{equation}
\item Recall: It answers the question that out of real positive examples, what proportion did the classifier labeled as positive? It is defined as the following formula:
\begin{equation}
	recall=\frac{TP}{TP+FN}
\end{equation}
\item F-measure: It is a measure which combines precision and recall and is defined according to the following formula:
\begin{equation}
	F=2\times{\frac{precision\times{recall}}{precision+recall}}
\end{equation}
\end{itemize}
 
\subsection{Analyzing Results}
In this section, we present the results based on the mentioned metrics in the previous section. The results for experiment 1 in which we did not ask the model to return the CWEs, are provided in \autoref{experiment1_table}. The precision for the model in this experiment is not better than other three tools. Furthermore, the low recall suggests that using this model for only detecting vulnerable lines of a code does not give any better results than SAST tools since low recall leads to high false negative rate. Likewise, the results of experiment 2, which are presented in \autoref{experiment2_table}, indicate that using GPT-3.5 model with all the classes given as labels, does not provide superior results in comparison with the SAST tools. Note that in this experiment, the order of the given labels to GPT-3.5 model has high impact in the generated results from the model. This is because when the order of the labels is changed, in fact the prompt is modified and as we mentioned before, the prompt has great effect on the model's results. Therefore, we gave the labels in a random order. Here, we reach the same conclusion as \cite{deeplearningg2} in which the authors concluded that the capabilities of the ChatGPT model for detecting vulnerabilities in code are limited. 

\begin{figure}
	\centering
	\includegraphics[width=0.5\textwidth]{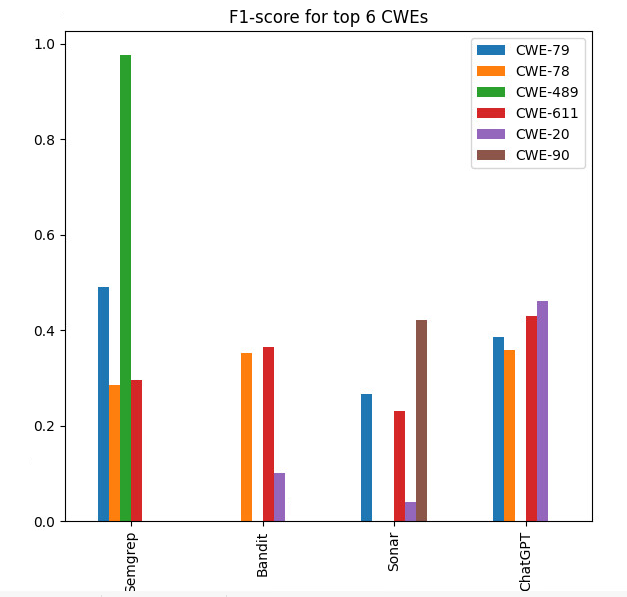} 
	\caption{F1-score of top 6 CWE classes in experiment 3 (case 2)}
	\label{F1_fig} 
\end{figure}

The results of experiment 3 in which we provided the classes per vulnerable file, are given in \autoref{experiment3_table}. Here the figures indicate that case 1, in which we do not accept the new labels returned from the model, has produced better results than case 2. These results are even significantly better than those of SAST tools in this experiment. The F1-score for the top 6 CWEs in terms of frequency are illustrated in \autoref{F1_fig} for this experiment. This behavior shows that using ChatGPT as an assistant along with SAST tools can be a good idea. Moreover, if we do not provide any labels for the model and ask it to return the CWEs of the vulnerable codes from its own knowledge, as we did in experiment 4, we obtain the results in \autoref{experiment4_table} which are comparable to the SAST tools. By and large, our experiments show that using ChatGPT model as an assistant for SAST tools can provide hopeful results.

\begin{table}
\tiny
\centering
    \begin{tabular}{|c|c|c|c|}
    \hline
	& \textbf{ Precision} & \textbf{ Recall} & \textbf{F1}  \\ \hline
	\textbf{Semgrep} & 0.6694 &\textbf{ 0.1504} & \textbf{0.2457} \\ \hline
	\textbf{Bandit} &  0.7450 & 0.1447 & 0.2424  \\ \hline
	\textbf{SonarQube} &  \textbf{0.9104} & 0.1161 & 0.2060  \\ \hline
	 \textbf{GPT-3.5} & 0.7413 &  0.0819 & 0.1475 \\ \hline
    \end{tabular}
    \caption{Results of Experiment 1 (Binary classification)}\label{experiment1_table}
 \end{table}


\begin{table}
\tiny
\centering
    \begin{tabular}{|c|c|c|c|}
    \hline
	& \textbf{ Precision} & \textbf{ Recall} & \textbf{F1}  \\ \hline
	\textbf{Semgrep} & \textbf{0.4682} & \textbf{0.1123} &  \textbf{0.1812} \\ \hline
	\textbf{Bandit}  & 0.3168 & 0.0609 & 0.1022   \\ \hline
	\textbf{SonarQube}  & 0.3283 & 0.0419 & 0.0743  \\ \hline
	 \textbf{GPT-3.5} & 0.1659 &  0.0761 & 0.1044 \\ \hline
    \end{tabular}
    \caption{Results of Experiment 2 (Selecting from the list)}\label{experiment2_table}
 \end{table}


\begin{table}
\tiny
\centering
    \begin{tabular}{|c|c|c|c|}
    \hline
	& \textbf{ Precision} & \textbf{ Recall} & \textbf{F1}  \\ \hline
	\textbf{Semgrep} 		& 0.4682 & 0.1123 &  0.1812 \\ \hline
	\textbf{Bandit} 		& 0.3168 & 0.0609 & 0.1022   \\ \hline
	\textbf{SonarQube}  & 0.3283 & 0.0419 & 0.0743  \\ \hline
	 \textbf{Experiment3,GPT-3.5-Case 1}  & \textbf{0.7807} & \textbf{0.2781} &\textbf{ 0.410}1 \\ \hline
	 \textbf{Experiment3,GPT-3.5-Case 2}  &0.333 & 0.1542 & 0.2109\\ \hline
	
    \end{tabular}
    \caption{Results of Experiment 3 (SAST assistant)}\label{experiment3_table}
 \end{table}


\begin{table}
\tiny
\centering
    \begin{tabular}{|c|c|c|c|}
    \hline
	& \textbf{ Precision} & \textbf{ Recall} & \textbf{F1}  \\ \hline
	\textbf{Semgrep} &  \textbf{0.4682} & 0.1123 &  \textbf{0.1812} \\ \hline
	\textbf{Bandit} & 0.3168 & 0.0609 & 0.1022   \\ \hline
	\textbf{SonarQube}  & 0.3283 & 0.0419 & 0.0743  \\ \hline
	 \textbf{GPT-3.5} & 0.3350 &  \textbf{0.1238} & 0.1808  \\ \hline
    \end{tabular}
    \caption{Results of Experiment 4 (Free Classification)}\label{experiment4_table}
 \end{table}


\section{Threats to Validity}\label{threats_section}
In this Section, we discuss some factors in our experiments that could affect the correctness of the results. Our biggest challenge was the choice of the prompts of ChatGPT. There are some metrics for measuring the effectiveness of a prompt for LLMs. In \cite{tony2023} naturalness and expressiveness are mentioned as two important factors for a prompt. Here, we tried to choose the most efficient prompts in terms of these metrics and based on what was explained in \ref{parameters_section}  \cite{chatgptcourse}. However, it is possible that a more careful selection of the prompt can affect the results. Another factor which may also affect the results is the size of the dataset and its accessibility on the Internet. Furthermore, the distribution of the CWEs of the dataset is of great importance. To overcome this threat, we chose three different datasets for better generalization of the vulnerabilities they cover, but there may be still few coverages of the vulnerabilities. Moreover, we only compare this model with three SAST tools for Python language. Perhaps, further SAST tools affect the results. And finally, we only test GPT-3.5 model of ChatGPT and it is possible that the new billable version (GPT-4) performs better than this version. 

\section{Conclusion}\label{sec:conclusion}
In this paper, we did four types of experiments with ChatGPT model to detect the security vulnerabilities of Python codes. We compared this model with Bandit, Semgrep and SonarQube that are popular SAST tools for Python codes. We concluded that using GPT-3.5 model for vulnerability detection of codes  in some especial manners  gives promising results. Specifically, if we use it as SAST tool assistant, it will produce results that can help to improve the returned results of SAST tools. Overall, we believe this model has the potential to be used in vulnerability detection tasks regarding the factors that may effect the correctness of the results which we described in \ref{threats_section}. However, we admit that this study is not general from all aspects and provides primary steps toward this path. In future studies, the behavior of the latest model of ChatGPT (GPT-4) which is more powerful than GPT-3.5 model, can be examined in vulnerability detection of codes with the hope of obtaining better results. Moreover, the Temperature  parameter of the model can be set to values other than zero and innovative rules can be passed to decide for the most efficient obtained results. Another suggestion can be using one-shot learning in future works. Moreover, it should be considered that, there is a security caution about using ChatGPT as as a SAST tool because it is required to upload the source code on the OpenAI servers.

\appendix
\section{Appendix} \label{sec:appendix}
\chapter{}\label{appendix:raw}
The distribution of labels of our dataset is provided in \autoref{dataset_table}


\begin{table}[htpb]
\caption{Details of Datasets}
\centering
\SetTblrInner{rowsep=-2pt}
\begin{tblr}{|c|c|c|}
	\hline
      Vulnerability  & Occurrence in \cite{siddiq}  & Occurrence in \cite{pyt1} \\
	\hline
	CWE-15 & 15 & 0 \\
	CWE-1004& 1 & 0 \\
	CWE-614& 1 & 0 \\
	CWE-489& 22 & 0 \\
	CWE-20& 0 & 18 \\
	CWE-22& 3 & 11 \\
	CWE-78& 25 & 5 \\
	CWE-79& 26 & 11 \\
	CWE-80& 0 & 3 \\
	CWE-89& 2 & 6 \\
	CWE-90& 0 & 15 \\
	CWE-94& 0 & 7 \\
	CWE-95& 0 & 2 \\
	CWE-99& 0 & 2 \\
	CWE-113& 0 & 5 \\
	CWE-116& 0 & 7 \\
	CWE-117& 0 & 6 \\
	CWE-1204& 0 & 3 \\
	CWE-193& 0 & 4 \\
	CWE-200& 0 & 5 \\
	CWE-209& 0 &  4 \\
	CWE-215& 0 & 4 \\
	CWE-250& 0 & 3 \\
	CWE-252& 0 & 2 \\
	CWE-259& 0 & 2 \\
	CWE-269& 0 & 4 \\
	CWE-283& 0 & 2 \\
	CWE-284& 0 & 3 \\
	CWE-285& 0 & 5 \\
	CWE-295& 1 & 8 \\
	CWE-297& 0 & 10 \\
	CWE-306& 0 & 4 \\
	CWE-312& 0 & 3 \\
	CWE-319& 0 & 2 \\
	CWE-321& 0 & 1 \\
	CWE-326& 0 & 4 \\
	CWE-327& 0 & 7 \\
	CWE-329& 0 & 4 \\
	CWE-330& 0 & 1 \\
	CWE-331& 0 & 1 \\
	CWE-339& 0 & 4 \\
	CWE-347& 0 & 3 \\
	CWE-352& 0 & 1 \\
	CWE-367& 0 & 3 \\
	CWE-377& 0 & 3 \\
	CWE-379& 0 & 4 \\
	CWE-384& 0 & 4 \\
	CWE-385& 0 & 6 \\
	CWE-400& 0 & 3 \\
	CWE-406& 0 & 9 \\
	CWE-414& 0 & 7 \\
	CWE-425& 0 & 4 \\
	CWE-434& 0 & 9 \\
	CWE-454& 0 & 6 \\
	CWE-462& 0 & 5 \\
	CWE-477& 0 & 2 \\
	CWE-488& 0 & 4 \\
	CWE-502& 0 & 15 \\
	CWE-521& 0 & 5 \\
	CWE-522& 0 & 12 \\
	CWE-595& 0 & 4 \\
	CWE-601& 0 & 14 \\
	CWE-605& 0 & 4 \\
	CWE-611& 0 & 22 \\
	CWE-641& 0 & 3 \\
	CWE-643& 0 & 8 \\
	CWE-703& 0 & 13 \\
	CWE-730& 0 & 10 \\
	CWE-732& 0 & 4 \\
	CWE-759& 0 & 4 \\
	CWE-760& 0 & 2 \\
	CWE-776& 0 & 3 \\
	CWE-798& 0 & 4 \\
	CWE-827& 0 & 3 \\
	CWE-835& 0 & 5 \\
	CWE-841& 0 & 10 \\
	CWE-918& 0 & 8 \\
	CWE-941& 0 & 8 \\
	CWE-943& 0 & 7 \\
	\hline
\end{tblr}
\label{dataset_table}
\end{table}

%

\bibliographystyle{unsrt}
\bibliography{biblio}

\noindent
\\

\begin{wrapfigure}[8]{l}{1.5cm}
	\begin{center}
		\includegraphics[width=2.3cm, height= 2.85cm]{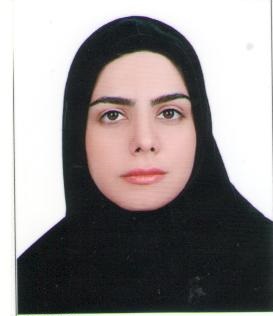}
	\end{center}
\end{wrapfigure}
\noindent
\newline
{\bf Atieh Bakhshandeh} {is a cyber security researcher in RCDAT since 2014. She has a master degree in Computer Science and her interested research areas include data analysis for security threat detection and penetration testing.}

\noindent
\newline
\begin{wrapfigure}[8]{l}{1.5cm}
	\begin{center}
		\includegraphics[width=2.3cm, height= 2.85cm]{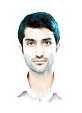}
	\end{center}
\end{wrapfigure}
\noindent
\newline
{\bf  Abdalsamad Keramatfar} { received his PhD in Information Technology engineering in 2021 with a focus on Natural Language Processing and Deep Learning. He worked as a Data Scientist for 6 years at SID and is currently working as an AI researcher at RCDAT.}


\begin{wrapfigure}[8]{l}{1.5cm}
	\begin{center}
		\includegraphics[width=2.3cm, height= 2.85cm]{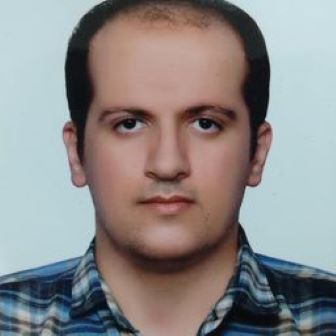}
	\end{center}
\end{wrapfigure}
\noindent
\newline
{\bf Amir Norouzi} { is a data scientist with a Master's degree in Bioelectric Engineering from Amirkabir University of Technology. He is experienced in machine learning, data engineering, deep learning, and data analysis. He has been working in AI since 2017 and is currently working as a researcher in RCDAT.}

\begin{wrapfigure}[8]{l}{1.5cm}
	\begin{center}
		\includegraphics[width=2.3cm, height= 2.85cm]{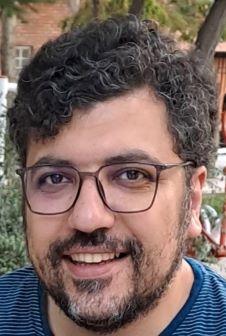}
	\end{center}
\end{wrapfigure}
\noindent
\newline
{\bf Mohammad Mahdi Chekidehkhoun} {graduated from Telecommunications engineering in 2016 with a focus on identifying threats in the mobile network. He has been working as a security specialist at RCDAT for 12 years, focusing on penetration testing and code security reviews.}

\end{document}